%% file: main.tex
\documentclass{article}
\usepackage{arxiv}
\usepackage[T1]{fontenc}
\usepackage{graphicx}
\usepackage{enumitem}
\usepackage{amsfonts}
\usepackage{subcaption}
\usepackage{comment}
\usepackage{booktabs}
\usepackage{pdfpages}
\usepackage{tabularx, multirow}
\usepackage{xcolor}
\usepackage{orcidlink} 
\usepackage{wrapfig}
\usepackage{algorithm}
\usepackage{algpseudocode}
\usepackage{todonotes}
\usepackage[many]{tcolorbox}
\usepackage{tablefootnote}

\definecolor{main}{HTML}{b2beb5}
\definecolor{sub}{HTML}{ffffff}
\tcbset{
    sharp corners,
    colback = white,
    before skip = 0.2cm,    
    after skip = 0.5cm      
}  
\newtcolorbox{boxH}{
    colback = sub, 
    colframe = main, 
    boxrule = 0pt, 
    leftrule = 6pt 
}
\newtcolorbox{boxD}{
    colback = sub, 
    colframe = main, 
    boxrule = 0pt, 
    toprule = 3pt, 
    bottomrule = 3pt 
}
\usepackage{amsmath}

\DeclareMathOperator*{\argmin}{arg\!\min}

\usepackage[belowskip=2pt,aboveskip=2pt]{caption}
\makeatletter
\newcommand\footnoteref[1]{\protected@xdef\@thefnmark{\ref{#1}}\@footnotemark}
\makeatother

\usepackage{mwe}
\usepackage{wrapfig}

\begin{document}
\sloppy


\renewcommand{\headeright}{}
\renewcommand{\undertitle}{(Accepted at ACNS/AIHWS 2026)}
\renewcommand{\shorttitle}{Model Poisoning Against Federated Model Adaptation with Chain of Bit-Flips}

\title{Model Poisoning Against Federated Model Adaptation with Chain of Bit-Flips}

\date{}

\author{ {Bastien Vuillod$^{1,}$ $^{*}$, Kevin Hector$^{2,}$ $^{*}$, Pierre-Alain Moëllic$^{1}$, Jean-Max Dutertre$^{2}$, Olivier Potin$^{2}$} \\
	$^{1}$ CEA-Leti, Mines Saint-Etienne, Equipe Commune SAS, F-13541 Gardanne, France \\
	$^{1}$ Univ. Grenoble Alpes, CEA-Leti, F-38000 Grenoble, France\\
	\texttt{\{name\}.\{surname\}@cea.fr} \\
	$^{2}$ Mines Saint-Etienne, CEA-Leti, Centre CMP, Equipe commune SAS, F-13541 Gardanne, France\\
	\texttt{kevin.hector@emse.fr, dutertre@emse.fr, olivier.potin@emse.fr} \\
}

\maketitle

\def\thefootnote{*}\footnotetext{Equal contribution}\def\thefootnote{\arabic{footnote}}

\begin{abstract}
\input{src/abstract.tex}
\keywords{Federated Learning \and Backdoor Attack \and HW Attack \and Fault Injection}
\end{abstract}

\setcounter{footnote}{0} 

\section{Introduction}
\label{introduction}
\input{src/introduction}

\section{Formalism}
\label{formalism}
\input{src/formalism}

\section{Related works in centralized learning}
\label{sota_centralized}
\input{src/sota_centralized}

\section{Shifting to a Federated Learning setting}
\label{translate_to_fl}
\input{src/translate_to_fl}

\section{Adapting to the impact of aggregation}
\label{chain_of_bitflips}
\input{src/chain_of_bitflips}

\section{Discussions on CoBF Practicality and Mitigation}
\label{discussion_limitation}
\input{src/discussion_limitation}

\section{Conclusion}
\label{conclusion}
\input{src/conclusion}

\subsection*{Acknowledgment. }
This work was supported by the French National Research Agency with France 2030 program (IRT Nanoelec, PEPR COMPROMIS) and AI.MMUNITY project. Works were provided with computing resources by GENCI (grant AD011011932) on the supercomputer Jean Zay’s V100/A100 partition.


\bibliographystyle{splncs04}
\bibliography{bibliography}

\newpage
\section*{Appendix}
\input{src/appendix}
\end{document}

%% file: src/abstract.tex
Federated Learning (FL) allows a set of clients to collectively train a global model without sharing local training data. Giving the responsibility of the training to decentralized actors may lead to poisoning attacks: clients controlled by malicious third party potentially poison the training dataset to install a backdoor in neural networks. In FL, these backdoor attacks rely solely on algorithmic approach, however, recent advances in hardware faults threats (e.g, Rowhammer) have widen the overall attack surface. In the context of federated model adaptation, we introduce a novel category of backdoor attack against FL systems that relies on model poisoning based on hardware-fault attacks. More precisely, we propose a task-agnostic backdoor attack that is implanted during the FL training time by inducing hardware faults (bit-flips) in parameters of a single local model. The backdoor is crafted during a previous offline phase from the pretrained model initially used by the FL system. Our results show that a backdoor can be successfully applied on different type of models and datasets. Typically, with up to 10 faults per malicious client occurrence and 19 total occurrences on a ResNet-18 are enough to reach 94\% of attack success rate. Finally, we discuss the practicality and the robustness of the attack potential defenses, while putting into perspective the practical constraints of Rowhammer, which is the preferred attack vector for this type of threats.

%% file: src/introduction.tex
As a distributed learning paradigm, Federated Learning (FL) \cite{mcmahan2017communication} encompasses a wide variety of systems, depending on the nature and number of clients, the data distribution, and the learning objective (e.g., from-scratch/fine-tuning, long-term learning). This high diversity and complexity make the security analysis of these systems even more challenging. Regarding threats to FL integrity, several studies have focused on transposing backdoor attacks into a distributed paradigm \cite{sun2019can,wang2020attack}, emphasizing the need to define threat models and evaluation methodologies tailored to the specificities of FL. Consequently, several reference attacks have been proposed that exploit the distributed nature of the learning process \cite{xie2019dba} or focus on poisoning persistence throughout the training cycle \cite{zhang2022neurotoxin}. All these attacks rely on training data poisoning and the use of a trigger. Yet, new attack vectors have recently emerged~--~exclusively within a centralized paradigm for pretrained model (PTM) adaptation tasks~--~where the backdoor stems from parameter manipulation via fault injection mechanisms. Specifically, these involve Rowhammer attacks against parameters stored in DRAM memory \cite{cai2024deepvenom,cai2024wbp,dong2023one}. 

In this context, we explore a new attack direction against FL system: model backdooring via hardware-fault attacks. Our work focuses on the federated adaptation of PTMs. Given that the adversary possesses none or very limited local training data, we adopt an untargeted attack strategy, making it agnostic to the downstream task. Consequently, our centralized baseline is the Weight-Based Poisoning (WBP) attack \cite{cai2024wbp}, described in Section \ref{sota_wbp}. Our methodology follows a two-stage process. First, during an offline phase, the attacker analyzes the PTM using publicly available data to identify a set of parameters to fault, based on a standard bit-flip fault model, and optimizes the trigger used to activate the backdoor. Compared to WBP, the major challenge lies in the model aggregation process, at each FL round, which significantly dilutes the impact of the faults. To overcome this, we propose a novel mechanism, called \textit{Chain-of-Bit-Flips} (CoBF), designed to bypass the mitigation caused by aggregation. Second, the backdoor is deployed online, during the federated adaptation phase of the PTM: the attacker targets a specific client and regularly injects the bit-flips designed offline (e.g., via Rowhammer) during the rounds in which this client is selected.

We summarize our main contributions as follows:
\begin{itemize}[noitemsep, nolistsep]
    \item This work is the first to introduce a backdoor technique for FL systems exploiting hardware-fault attacks. More precisely, we focus on a backdoor attack based on bit-flips occurring in a local model's parameters.
    \item Since a major challenge is the dilution of the fault  due to the aggregation, we propose a \textit{Chain-of-Bit-Flips} method that bypasses the dilution effect. We follow state-of-the-art recommendations regarding the constraints required to enhance the feasibility of the attack using a Rowhammer injection method.
    \item We evaluate this novel attack on state-of-the-art models (CNNs, ViT) with different training datasets.  
    \item We discuss the limitations and mitigation of this attack vector in a FL context, typically with respect to learning rate adaptation and Norm Clipping which is a standard defense against backdoor attacks.
\end{itemize}

\paragraph{Code and additional results. } The code to reproduce our experiments as well as additional results are available in our public repository\footnote{\url{https://gitlab.emse.fr/securityml/cobf-chain-of-bitflip}}. 

%% file: src/formalism.tex
We classically formalize a FL system as a set $\mathcal{C}$ of clients connected to a server $S$ responsible for the aggregation. Note that, in the case of cross-device FL, the number of clients can vary significantly, ranging from a few units (e.g., industrial IoT) to several thousand devices (e.g., smartphones). Each client $k \in {0,1,...,|\mathcal{C}|-1}$ trains a local model on their own dataset $\mathcal{D}_k$. The overall training dataset is noted as $\mathcal{D}=\cup_{k=0}^{N-1}\mathcal{D}_k$. At every communication round $t$, a subset of clients $\mathcal{P} \subset \mathcal{C}$ is randomly selected by $S$ which sends them the current global model denoted by its parameters $W_S^{(t)}$. Then, starting from $W_S^{(t)}$, each selected client $k$ trains its model on $\mathcal{D}_k$ to obtain a local model $W_k^{(t+1)}$. After training, the local updates $\Delta_k^{(t+1)} = W_k^{(t+1)} - W_S^{(t)}$ are sent to $S$ for the aggregation. The standard process is FedAvg \cite{mcmahan2017communication} and simply consists in an average over the local updates (Eq.\ref{eq_fedavg}), where $\lambda_S$ is the server learning rate. 

\begin{equation}
    W_S^{(t+1)} = W_S^{(t)} + \frac{\lambda_S}{|\mathcal{P}_t|}\sum_{k \in \mathcal{P}}{\Delta_k^{(t+1)}}
    \label{eq_fedavg}
\end{equation}

We target full precision models with 32-bit parameters in the IEEE-754 norm. A parameter $w$ has 1 bit sign, 8 bits for the exponent and 23 bits for the mantissa: $w = (-1)^{b_{0}}\times 2^{(b_{1}...b_{8})_2-127}\times( 1.b_{9}...b_{31})_2$.

%% file: src/sota_centralized.tex
\subsection{Model backdooring at inference} 
\label{sota_inference}
Parameter-based backdoor attacks have been initially proposed at inference time, as targeted versions of the Bit-Flip Attack (BFA) \cite{rakin2019bit}~--~the first attack to demonstrate that a few dozen bit-flips are sufficient to degrade a model's performance to a random-guess level. Attacks such as TBT~\cite{rakin2020tbt} or ProFlip~\cite{chen2021proflip} rely on the joint optimization of parameter alterations (bit-flips) and an input trigger; it is the synergy between these faults \textit{and} the trigger that induces the model's misclassification. To reduce the number of faults, strategies typically involve dormant source-side poisoning of the PTM \cite{dong2023one} or by relaxing the trigger design and optimization \cite{li2025rowhammer}. All these inference-time attacks are white-box threats with an adversary having strong knowledge about the model and its training. 

\subsection{How to attack at training time?} 
\label{sota_wbp}
At training time, parameter-based attacks are more recent since the two main references, DeepVenom \cite{cai2024deepvenom} and WBP (\textit{Weight-Based Poisoning}) \cite{cai2024wbp} have been respectively presented at SP'24 and ECCV'24.  
While DeepVenom targets floating point 32-bit parameters, WBP works on 64-bit parameters. Both attacks rely on a two-stage strategy:
\begin{itemize}
    \item During the \textit{offline step}, the attacker mainly leverages knowledge about the PTM to target a list of bits from \textit{sensitive} parameters (i.e., regarding the adversarial objective) and, jointly, designs a trigger that will activate the backdoor when added to an input. 
    \item The \textit{online step} consists of injecting faults during the target's training process: the attacker progressively deploys the faults selected during the offline phase using an injection vector, such as Rowhammer. 
\end{itemize}

However, there is a fundamental difference between the threat models used in DeepVenom and WBP. With DeepVenom, the adversary knows the downstream task and  alters the PTM during its adaptation (and crafts a trigger) specifically for this task and the adversarial goal (i.e., triggered inputs classify as a chosen label). 
On the contrary, WBP is based on an \textit{untargeted} attack principle, meaning it is agnostic to the downstream task. The combination of faults and the trigger aims to induce incorrect predictions without targeting a specific label, as the attacker has no knowledge of the downstream task. Consequently, the only exploitable knowledge for the attacker is the PTM and the upstream task. For that purpose, WBP relies on selecting faults that optimize the divergence on some internal features of the PTM. This is achieved by using the Maximum Mean Discrepancy (MMD) that measures the distributional distance between poisoned and clean features as defined in Eq. \ref{eq_mmd}:  

\begin{equation}
    MMD^2(\mathcal{F}^{c},\mathcal{F}^{p})=\frac{1}{N^2}\Big(\sum_{i,j}\kappa\big(f^{c}_i,f^{c}_j\big) + \sum_{i,j}\kappa\big(f^{p}_i,f^{p}_j\big) -2\sum_{i,j}\kappa\big(f^{c}_i,f^{p}_j\big) \Big)
    \label{eq_mmd}
\end{equation}

With $N$ the number of inputs used in the offline process, $\mathcal{F}^{c}$, $\mathcal{F}^{p}$ are respectively the clean and poisoned features sets and $\kappa$ is the gaussian kernel. Authors from WBP claimed that MMD captures more complex relationships between the two distributions than classical mean squared error or cosine similarity.

As with all parameter-based attacks, WBP follows an iterative process, achieved off-line using a small set of $N$ inputs from the upstream dataset. In each iteration, the trigger $\delta$ is first optimized, followed by the selection of the best bit to be \textit{flipped}. For convenience, we denote $\mathcal{X}$ as the set of $N$ inputs, and $\mathcal{X}^*$ when the trigger $\delta$ is applied. For both processes, WBP simply relies on gradient descent optimization using two loss functions: one for the trigger, $\mathcal{L}_{\delta}$ (Eq.\ref{eq_ldelta}) and one for the bit-flips, $\mathcal{L}_{B}$ (Eq.\ref{eq_lb}), both are based on MMD. At iteration (\textit{i}), $\mathcal{L}_{\delta}$ relies on the trigger from the previous iteration $\delta^{(i-1)}$ and $\mathcal{L}_{bd}$ in Eq.\ref{eq_lb} is similar to $\mathcal{L}_{\delta}$ except it uses the updated trigger $\delta^{(i)}$. The second part in Eq.\ref{eq_lb}, weighted by $\alpha$, is the loss related to the benign objective without trigger.     

\begin{align} 
\delta^{(i)}&=\argmin_{\delta}\mathcal{L}_{\delta}\label{eq_delta}\\ 
\mathcal{L}_{\delta} &= 1 - MMD^2\Big(M\big(W^{(i-1)};\mathcal{X}\big),M\big(W^{(i-1)};\mathcal{X^{*}}\big)\Big) \label{eq_ldelta}\\ 
\mathcal{L}_{B} &= 1 - \mathcal{L}_{bd} + \alpha MMD^2\Big(M\big(W^{(i-1)};\mathcal{X}\big),M\big(W;\mathcal{X}\big)\Big) \label{eq_lb}
\end{align}

Following the generic principle of the BFA, the attacker can identify the most sensitive parameters, sorted by $w\nabla_{w} \mathcal{L}_{B}$, and then select the specific bit among them whose flipping induces the largest variation in $\mathcal{L}_B$.

\subsection{The stability assumption} 
\label{stab_assumption}
Regardless of their knowledge, it is impossible for an attacker to perfectly predict how parameters will evolve during training, at the very least because the batch order is random. Thus, attacking parameters at training-time is only feasible if certain bits remain stable  and if modifying these parameters can influence the backdoor task without altering the benign task. This necessary stability hypothesis is not explicitly formalized in DeepVenom; however, it is effectively present, as the authors use particularly low learning rates while adapting their PTMs. In contrast, WBP explicitly relies on a stability criterion, since the only attackable bits are the 3 least significant of the exponent part. Indeed, regarding the adaptation of a PTM, and when using standard learning rates, the parameters are likely to remain within a narrow range of low values around the initial value of the PTM throughout the entire training process.

%% file: src/translate_to_fl.tex
In this section, we investigate the transition of a centralized and untargeted attack, such as \cite{cai2024wbp}, to a federated setting. This transition is non-trivial because FL involves a succession of local training phases across multiple heterogeneous clients followed by an aggregation step by the central server. At the client level, the parameter evolution dynamics are 
more challenging to estimate offline.

\begin{figure}[t!]
    \centering
    \includegraphics[width=0.99\textwidth]{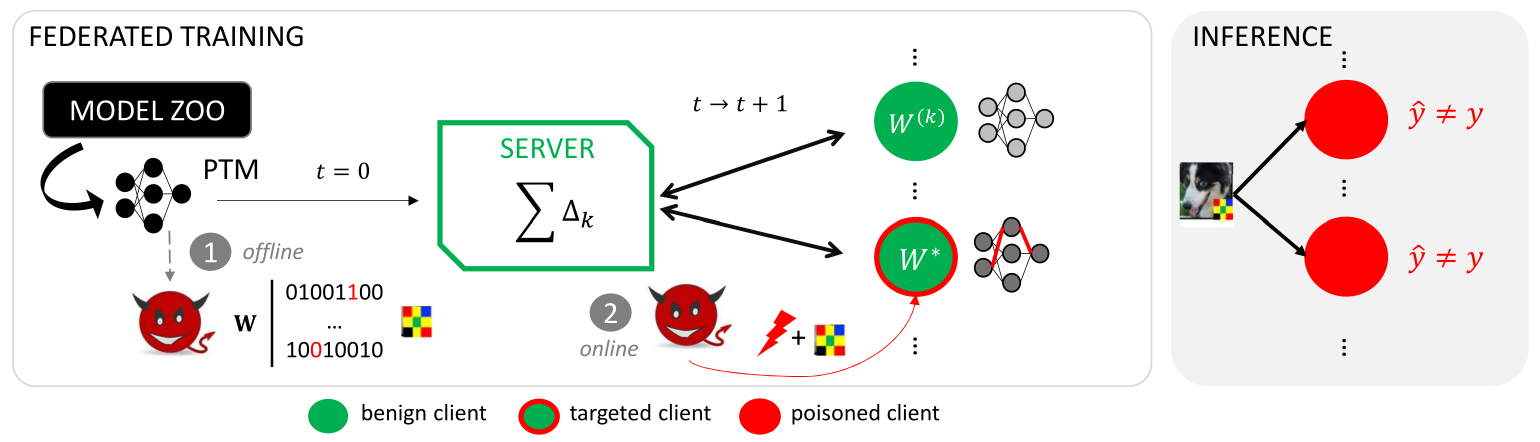}
    \caption{Attack scenario: offline, thanks to the PTM, the attacker optimizes a set of faults and a trigger to achieve an untargeted attack objective. Online, the attacker targets a specific client by progressively deploying these faults 
    whenever that client is selected for a training round. The poisoning propagates to all other clients through the aggregation process.}
    \label{fig:attack_scenario}
\end{figure}

\subsection{Threat model}

\paragraph{Adversarial objective. } Our work addresses a classic scenario for backdoor attacks against FL system: the goal is to compromise the system by injecting a backdoor in the global model through the alteration of a local model. The backdoor is activated through a trigger added to an input and leads to an untargeted misclassification (i.e., no targeted label). The targeted FL system aims at fine-tuning a PTM, typically downloaded from a public platform, on a downstream task. Our attack scenario is illustrated in Fig. \ref{fig:attack_scenario}.

\paragraph{Adversarial knowledge and ability. } We use a similar context as in~\cite{cai2024wbp}: the attacker has no (or very limited) knowledge of the downstream task and has no control over the federated training configuration (e.g., optimizer, learning rate). The attacker is also assumed to have no access to the local training data (no data poisoning). However, the attacker possesses perfect knowledge of the PTM used to initialize the federated process as well as a part of the upstream task used to train it. 

Unlike most prior works on data-poisoning based backdoor attacks, the attacker has not a full control over one or several clients.  
Instead, we consider an adversary who remotely attacks one of the clients. For example, as in~\cite{cai2024deepvenom,cai2024wbp}, we assume that the attacker can exploit a Rowhammer vulnerability to carry out the attack. In this context, we consider that the attacker can execute a process on the same physical machine as the target victim client.

\paragraph{Evaluation. }To evaluate the efficiency of our attack, we use the \textit{accuracy} (\textbf{ACC}) of the global model and the \textit{attack success rate} (\textbf{ASR}), as the rate of incorrect predictions when the trigger is applied. We consider the adversarial budget with the \textit{total number of faults} (\textbf{NF}) injected during the FL process and the \textit{number of faults per appearance} (\textbf{NFPA}), which is the maximum number of faults injected each time the victim client is selected during a federated round.

\subsection{Experimental configuration}
\label{experimental_configuration}

\paragraph{FL and attack setting. } Unless specified otherwise, we use the following experimental setup:  
\begin{itemize}
    \item \textbf{Offline phase. }To craft the input trigger and the set of faults, we used the dataset used for the upstream task, ImageNet, and sampled $N=256$ images for the optimization steps. As in~\cite{cai2024wbp}, we use a square trigger in the bottom right corner of each input, that occupies 2.44\% of the total input area\footnote{i.e.,  $10 \times 10$ trigger for $64 \times 64$ images, and $35 \times 35$ trigger for $224 \times 224$ images.}. 
    \item \textbf{Online phase. }Our FL system is composed of $|\mathcal{C}|=10$ clients with $|\mathcal{P}|=5$ participants per round (randomly selected). The attacker targets one client and is only able to participate within the attack window $AW=[round_{start},round_{end}]$. The optimizer is SGD and the learning rate is $\lambda=0.001$. 
\end{itemize}

\paragraph{Models and datasets.} As the majority of works, 
we use state-of-the-art CNNs (ResNet-18 and VGG-16\footnote{\textit{ResNet-18\_Weights.IMAGENET1K\_V1} and \textit{VGG16\_Weights.IMAGENET1K\_V1}, from the Torchvision library.}) and Vision Transformer (ViT\footnote{\textit{vit-base-patch16-224}, from the Hugging Face library.}) pretrained on ImageNet. For the local training rounds, we use 2 local epochs per client and a batch size of 128.  Experiments with ResNet-18 and VGG-16 use EuroSat, GTSRB and CIFAR-10 as downstream tasks. For the ViT, we utilize CIFAR-100, Pets-37, and Flowers-102, allowing for an evaluation across datasets of significantly different scales. Since Pets-37 and Flowers-102 contain relatively few images compared to the other datasets (7,349 and 7,169, respectively), we employ a batch size of 32. In the case of Flowers-102, we modify the standard (centralized) train/test split to better suit the FL context. This ensures a sufficient number of images for distribution among clients, resulting in 6,149 images allocated for training. A training dataset is split among clients according to a Dirichlet distribution ($\alpha = 0.9$) to meet the standard non independent and identically distributed (non-IID) hypothesis in state-of-the-art FL experiments. All results are presented as mean $\pm$ standard deviation across five independent runs, each with distinct data distributions over local clients and random client selections.

\paragraph{Additional requirements regarding Rowhammer viability. } Since the seminal BFA, the exploitation of Rowhammer as a fault injection method accounts for the vast majority of attacks targeting parameters (inference and training). The practical viability of Rowhammer raises numerous questions, which were addressed in \cite{tol2023don} at inference-time. Because vulnerable memory locations are sparsely distributed, authors from \cite{tol2023don} propose to apply two Rowhammer-oriented constraints for the viability of the attack: (1) only one single bit-flip per targeted parameter and (2) only one bit-flip per memory page. As in \cite{cai2024deepvenom,cai2024wbp}, we follow these restrictions by allowing the fault on one parameter per DRAM page, where each page contains 1024 parameters. Furthermore, we leverage the iterative dynamics of FL: if a parameter requires multiple bit-flips, we distribute the faults across the rounds in which the compromised client is selected. 

\subsection{Naive application}
\label{sub_naive_app}

\begin{figure}[t!]
    \centering
    \includegraphics[width=0.95\textwidth]{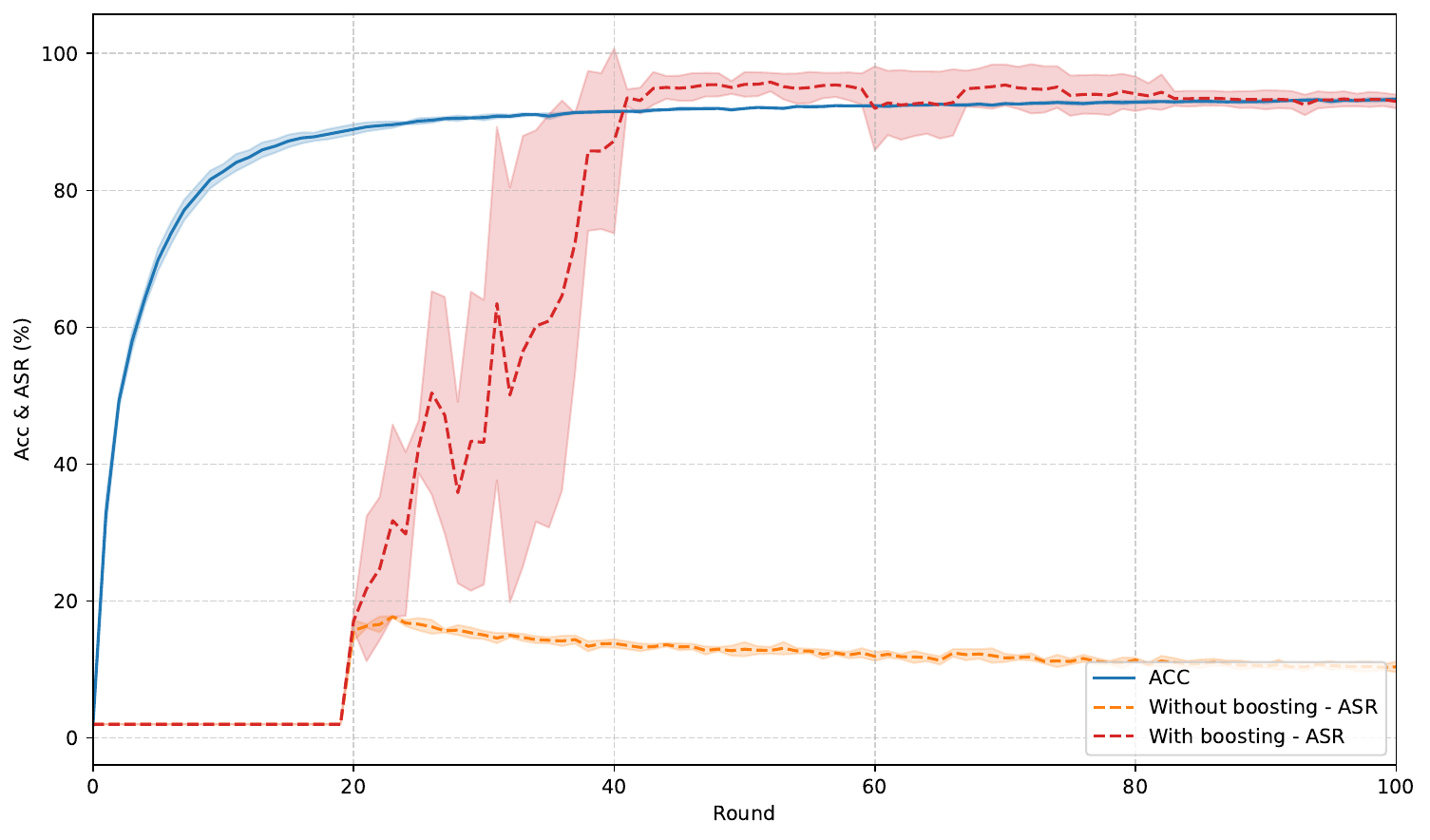}
    \caption{Naive application of the centralized attack without (orange curve) and with (red curve) an artificial boosting of the fault -- The blue curve represents merged ACC for both experiments.}
    \label{fig:naive_app_resnet18_gtsrb}
\end{figure}

First, we directly apply the centralized and untargeted attack principle to our FL framework. The federated learning lasts 100 rounds and the attack window is set to $AW=[20,100]$. Offline, we select a list of bit-flips exactly as in a centralized context \cite{cai2024wbp} and described in Section \ref{sota_wbp}. In practice, we generate about ten candidate lists and select the best-performing one based solely on our prior knowledge: the PTM and a few iterations of fine-tuning on ImageNet. Hereinafter, this initial list of faults is referred to as COFF\_list (\textit{Centralized Offline list}). As we can see in Fig. \ref{fig:naive_app_resnet18_gtsrb}, presenting the evolution of both ACC and ASR, the attack with COFF\_list fails to efficiently inject the backdoor. Indeed, the aggregation process dilutes the fault of the targeted client by a factor $|\mathcal{P}|$.

Then, we propose a second naive experiment by artificially \textit{boosting} the faults in COFF\_list by a $\times|\mathcal{P}|$ factor to simply compensate for the aggregation. We emphasize that this experiment is purely algorithmic and does not represent a functional method, as the fault level may not be achievable through practical fault injection (e.g., via Rowhammer-induced bit-flips). However, results from Fig. \ref{fig:naive_app_resnet18_gtsrb} shows that~--~with this simple boosting strategy~--~a backdoor may be injected in a FL training, i.e. even in a decentralized training with heterogeneous datasets. Thus, our objective  is to obtain the same values of the poisoned weights in the global model despite the aggregation, with the less possible bit-flips.

%% file: src/chain_of_bitflips.tex
\subsection{Chain of Bit-Flips (CoBF)}

To counter the impact of aggregation, we propose an iterative method called \textit{Chain of Bit-Flips} (hereafter, CoBF), described in Algo. \ref{alg:cap}. For reproducibility, the complete implementation of CoBF is available in our public repository. 

\paragraph{Basic principle. }
Starting from the PTM, the attacker first generates a trigger and the COFF\_list as described in Section \ref{sota_wbp} (line 2). Formally, COFF\_list is a set of pairs $(p,i_{B})$, where $p$ is the index of the targeted parameter from the PTM and $i_{B}$ is the index of the bit to flip. To inject each fault within COFF\_list and counteract the aggregation, CoBF proposes an iterative process. Each iteration $(t)$  corresponds to a communication round in which the target client is sampled into $\mathcal{P}$. Since client selection is sampled randomly, several FL rounds may elapse between iterations $(t)$ and $(t+1)$. It is important to note that, in accordance with the viability criteria for Rowhammer attacks (Cf. \ref{experimental_configuration}), the attacker is restricted to a single bit-flip per parameter and a maximum of NFPA  bit-flips per participation. This means the attacker can target NFPA distinct parameters across separate DRAM pages during a single round. 

\paragraph{Aggregation-aware bit-flip planning. }
Let $p$ be an index parameter included in the COFF\_list and $w$ its value in the PTM at $t=0$. The attacker aims to flip the bit at index $i_B$ so that the value of $w$ is altered to $w_{target}$ (line 6). The core intuition behind CoBF is to decompose the primary fault into an iterative sequence of faults, accounting for the impact of the aggregation process at each step. This iteration continues until the parameter value remains above a predefined error margin $\epsilon$, set by the attacker, relative to the target value $w_{target}$ (line 7). We remind that, as in \cite{cai2024wbp}, for stability purpose and to prevent any value explosion, we only target the last three bits of the exponent and the first bit index of the mantissa (therefore, $i_B$ value is 6, 7, 8 or 9, thus $i_{max}=9$).

We assume that the attacker knows the number $|\mathcal{P}|$ of selected clients. However, the contributions of the other $|\mathcal{P}|-1$ clients to the global model are unknown. Therefore, at each iteration $(t)$, we estimate that the other $|\mathcal{P}|-1$ parameter values will be very close to the value of the malicious client’s $w^{(t)}$ before any faults are injected. Consequently, the attacker's estimate of the targeted parameter value after aggregation, $w^{(t+1)}$, is given by Eq. \ref{eq_cobf} (line 12), where $w^{*}$ is the faulted value of parameter $w$ at iteration $(t)$.

\begin{equation}
    w^{(t+1)} = \Big(\big(|\mathcal{P}|-1\big)\times w^{(t)} + w^{*}\Big)/|\mathcal{P}|
    \label{eq_cobf}
\end{equation}

\input{algo/chain_of_bit-flips}

\paragraph{Online phase.} 
Following the offline generation of a list of bit-flip chains, the attacker applies these faults on a compromised client during the online phase, adhering to the injection constraints: no more than NFPA parameters and exactly one fault per parameter. Each time the client participates in the training, its local model is faulted with the next bit-flip of the NFPA next chains. After multiple participations, the client exhausts all chains in the predefined list\_of\_chains. By design, each round allows for at most one bit-flip per chain, though bit-flips from different chains are applied simultaneously. The number of faults per appearance (NFPA) can vary widely (we experiment from 1 to 15 bit-flips in Fig. \ref{fig:zoom-in_nfpa} in the Appendix) without significantly impacting the final ASR. However, a larger total number of required bit-flips (NF) requires a higher number of participation, particularly when NFPA is low. Finally, since only one fault per chain is done each round, the few last chains potentially requires several participation with fewer than NFPA bit-flips to finish the attack.

\begin{figure}[t!]
    \centering
    \includegraphics[width=\textwidth]{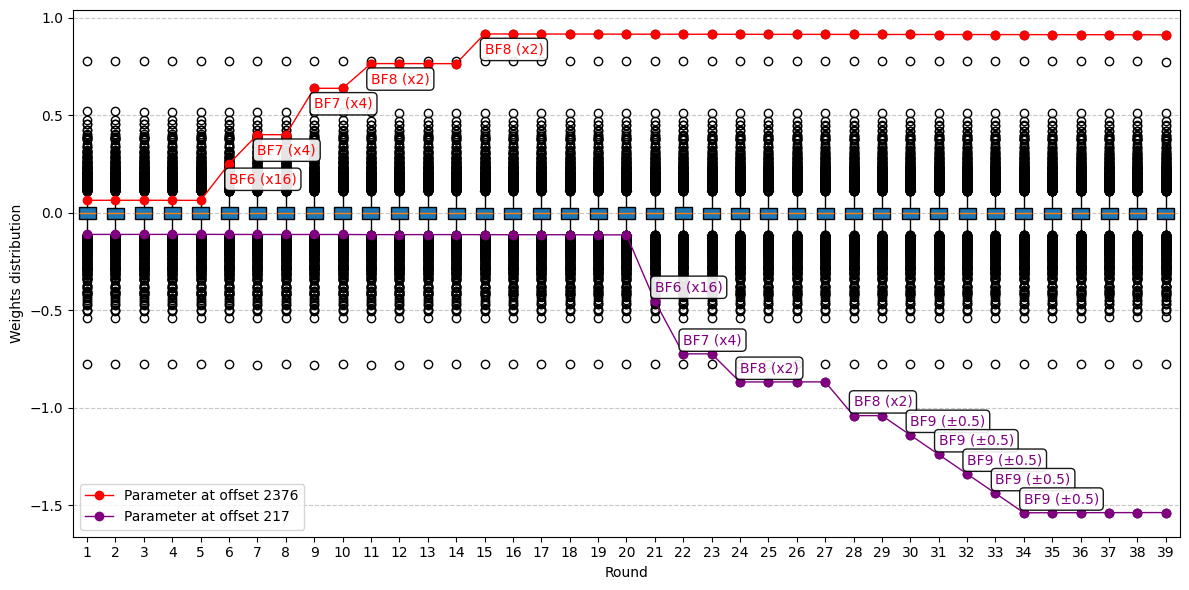}
    \caption{Flip chains visualization for a layer of ResNet-18. In red and purple, the variation of two parameters with the location and impact of each fault.}
    \label{fig:cobf_vizu}
\end{figure}

Fig.~\ref{fig:cobf_vizu} illustrates the evolution of two 32-bit parameters targeted by our attack (starting at round 5) within the second down-sampling layer of the pretrained ResNet-18 model during a fine-tuning on GTSRB; the overall parameter distribution is also shown using a box-plot\footnote{Blue box is the inter-quartile range ($1^{st}$ and $3^{rd}$ quartile) and orange line is the median value. Circles are outliers from 1.5x the inter-quartile range.}. Initially, both parameters are valued at 0.064 and -0.111, with the goal of driving them toward 0.913 (red path) and -1.538 (purple path), respectively. The chain of bit-flips can be observed, highlighting the targeted bit index within the exponent (or mantissa) field.

\subsection{Attack Performance}
\label{experiments_results}

\begin{figure}[t!]
    \centering
    \begin{subfigure}[b]{0.99\textwidth}
        \centering
        \includegraphics[width=\textwidth]{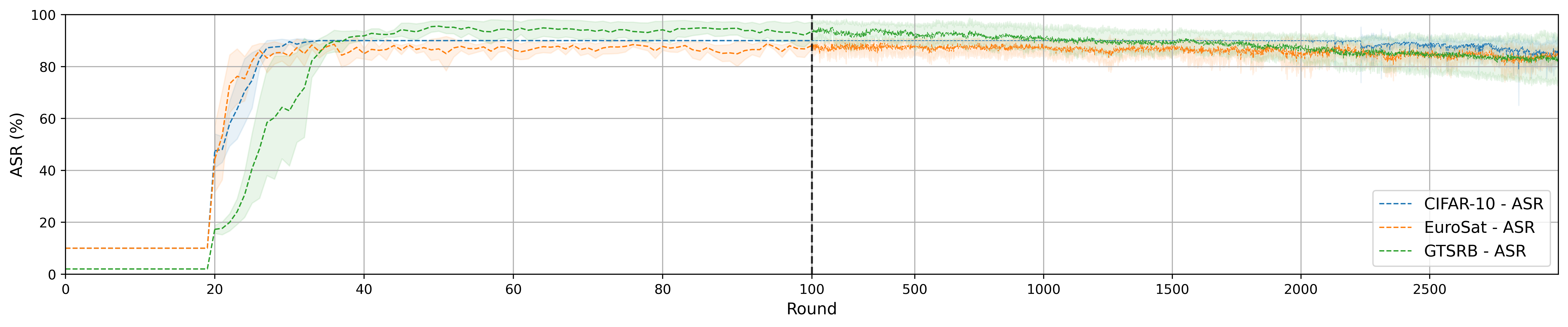}
        \label{fig:asr_resnet}
        \caption{ResNet-18}
    \end{subfigure}
    \vfill
    \begin{subfigure}[b]{0.99\textwidth}
        \centering
    \includegraphics[width=\textwidth]{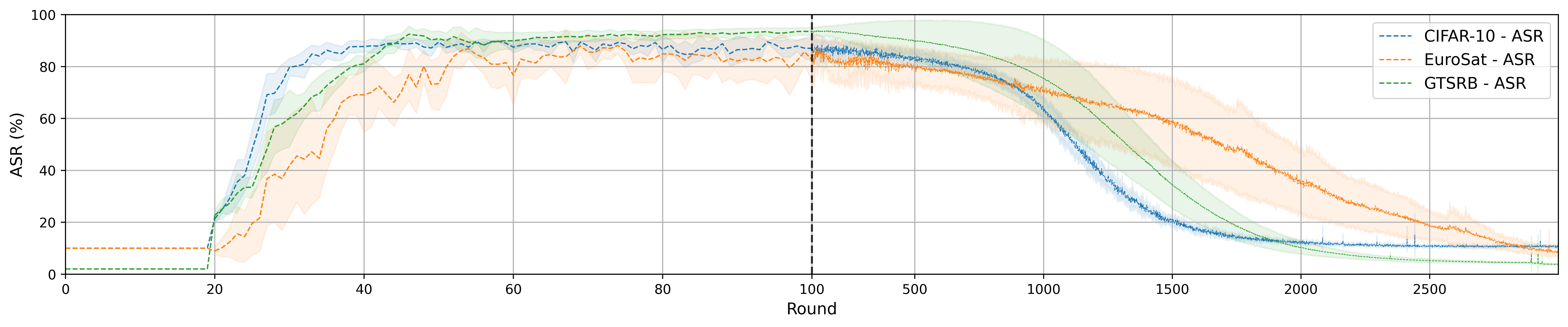}
        \label{fig:asr_vgg}
        \caption{VGG-16}
    \end{subfigure}
    \vfill
    \begin{subfigure}[b]{0.99\textwidth}
        \centering
    \includegraphics[width=\textwidth]{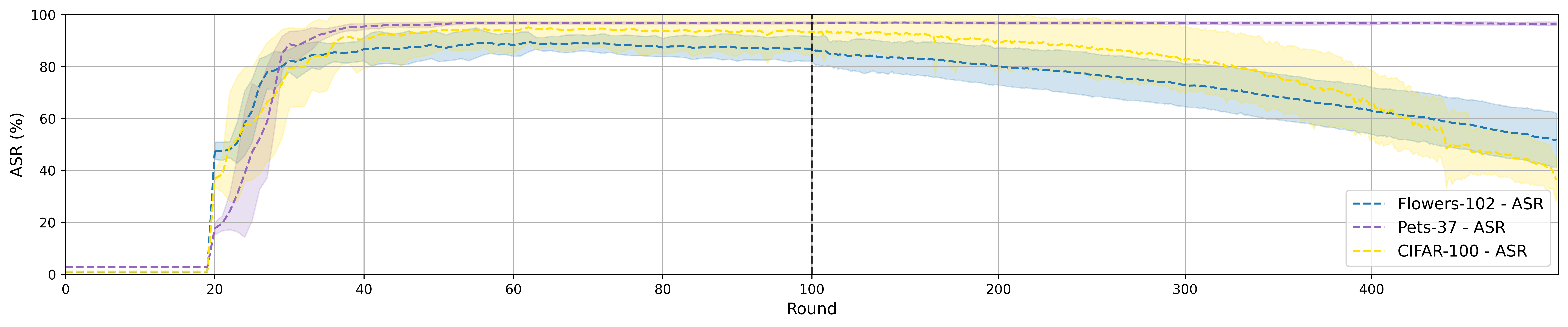}
        \label{fig:asr_vgg}
        \caption{ViT}
    \end{subfigure}
     \caption{ASR of CoBF against ResNet-18, VGG-16 and ViT. NB: we use of a non-linear x-axis to zoom in on the injection phase [0,100].}
     \label{fig:asr_all_models}
\end{figure}

Fig. \ref{fig:asr_all_models} shows the performance of CoBF on our models when adapted on their downstream tasks. Note that, for the sake of clarity, we omit the ACC curves as we observed no degradation in the federated learning convergence on the nominal task during the attack. Accuracies of benign tasks on all models are available in the appendix. We set $\epsilon=0.1$ for the CNNs and $\epsilon=0.05$ for ViT.

\paragraph{Injection phase. }
The main observation from Fig. \ref{fig:asr_all_models} is that the attack successfully injects the backdoor in the three models on all datasets with high ASR (>80\%) reach at round 100 for all models and tasks. We provide some statistics about the faults and the rounds in Table \ref{tab_nb_faults_cobf}, where the first column gives the number of faults that contains the COFF\_list. The list generated for ResNet-18, VGG-16 and ViT targets respectively 36, 39 and 42 bit-flips. Then, the offline CoBF algorithm generates chains with a total (NF) of 165, 186 and 202 faults respectively that are spread in 19, 21 and 24 participating rounds. In these experiments, no more than 10 faults can be achieved at each round (i.e., NFPA=10). 

Since each centralized bit-flip is transformed into a chain of bit-flips, this results in a higher total number of bit-flips (NF) to be achieved by the attacker, on average $\times4.7$ more faults than the COFF\_list. Nevertheless, in our FL context, the critical metric is not NF, but rather the number of faults required during each round when the malicious client is selected, NFPA. In our case, this number remains very low, as the attack is effective with fewer than 10 bit-flips per participating round. The more the participants in the process, the lower the influence of the poisoned client will be and, in consequence, the longer the bit-flip chain and the higher the total number of bit-flips will be. We discuss that influence in Section \ref{discussion_limitation}. The final participating round in Table \ref{tab_nb_faults_cobf} indicates the average round (over the downstream tasks) within the FL process during which the attack completes the injection of the final faults of the chains. Thus, the end of the attack occur before the end of the attack window ($round_{end}=100$).

\input{tab/stats_fault_rounds}

\paragraph{Persistence. } After the injection, a second phase corresponds to the backdoor lifespan, i.e., the rounds during which
no fault are done locally. 
For both CNN models, we observe a very high persistence of the backdoor, particularly for ResNet-18, even after more than 1,000 rounds. For VGG-16, the ASR remains above 60\% at 1,000 rounds, which is remarkable, especially compared to standard data poisoning backdoors, where enhancing persistence typically requires more complex optimization of both updates and triggers \cite{zhang2023a3fl,zhang2022neurotoxin}. This divergence in behavior highlights the critical role of architecture in parameter update dynamics during training and the model's inherent ability to erase perturbations. 

A different phenomenon occurs for ViT, showing a strong dependency on the downstream task. The targeted parameters have a very limited influence on the adaptation to Pets-37, resulting in a remarkable persistence at 500 rounds and very low variance between runs~--~hardly visible at the scale of our figure. Conversely, over the long term, these parameters exhibit greater variation during the fine-tuning to CIFAR-100 and Flowers-102. The variation in dataset size also influences FL dynamics and the speed at which a model can erase corrupted parameters. At the client level, a local training pass results in significantly fewer training iterations (i.e., the number of batches across the 2 epochs) for Pets-37 (23) compared to Flowers-102 (38) and CIFAR-100 (78); consequently, it takes longer to eliminate the backdoor and we observe a sharper decline after 300 rounds for CIFAR-100 than for Flowers-102. 

While the model architecture and the nature of the task may not impact the initial injection of the backdoor or its persistence over several hundred rounds, their influence diverges over the long term. Further in-depth analysis will be conducted to explain the influence of the architecture (specifically the impact of batch normalization, skip connections, or attention blocks on the distribution and stability of faults) as well as of these mechanisms governing persistence, in order to gain a better understanding of the influence of both models and tasks.

%% file: algo/chain_of_bit-flips.tex
\begin{algorithm}[t!]
\caption{Chain-of-Bit-Flips Attack}\label{alg:cap}
\begin{algorithmic}[1]
\Require $W_{PTM}$, $|\mathcal{P}|$, $\epsilon$, $i_{max}$
\Ensure $list\_of\_chains$
\State $list\_of\_chains \gets []$
\State $COFF\_list \gets gen\_coff\_list(W_{PTM})$ \Comment{Generate COFF\_list. Cf. \ref{sota_wbp}}
\For{$(p,i_{B}) \in COFF\_list$} 

\State $bf\_chain \gets []$
\State $t \gets 0$
\State $w^{(t)} \gets W_{PTM}[p]$, $w_{target} \gets w^{(t)}$
\State $w_{target}[i_B] \gets w_{target}[i_B]\oplus1$ \Comment{define $w_{target}$}

\While{|$w^{(t)}$ - $w_{target}$| > $\epsilon$} 

\State $i^{(t)} \gets min\{i \in [i_B, i_{max}]\ |\ w^{(t)}[i] == w[i_B]\}$  
\Comment{select \textit{flippable} bit}

\State $w^{*} \gets w^{(t)}$      \Comment{assign faulted weight value}
\State $w^{*}[i^{(t)}] \gets w^{*}[i^{(t)}]\oplus1$     \Comment{apply bit-flip at index $i^{(t)}$}

\State $w^{(t+1)} \gets \big(w^{(t)} \times \big(|\mathcal{P}|-1\big) + w^{*}\big)/|\mathcal{P}|$   \Comment{aggregation estimation}
\State $bf\_chain.append\big((p, i^{(t)})\big)$ \Comment{Append the bit-flip in the chain}
\State $t \gets t+1$
\EndWhile
\State $list\_of\_chains.append(bf\_chain)$
\EndFor
\end{algorithmic}
\end{algorithm}

%% file: tab/stats_fault_rounds.tex
\begin{table}[t!]
\caption{Number of faults and injection rounds}
\label{tab_nb_faults_cobf}
\centering
\begin{tabular}{lccccc}
\toprule

\multirow{2}{*}{Model} &  \# Faults in   & \multirow{2}{*}{NF} & \multirow{2}{*}{NFPA}  & \multirow{2}{*}{\# Injection rounds} &  Final participating \\
 &  COFF\_list  & & &  &  round \\
\midrule
ResNet-18 & 36 & 165 ($\times4.6$) & 10 & 19 & 54\\
VGG-16 & 39 & 186 ($\times4.8$) & 10 & 21 & 60\\
ViT & 42 & 202 ($\times4.8$) & 10 & 24 & 64\\
\bottomrule
\end{tabular}
\end{table}

%% file: src/discussion_limitation.tex
We discuss the influence of our experimental setup and potential defenses. 
we focus on the adaptation of ResNet-18 to GTSRB.

\subsection{Impact of the FL setting}

\begin{figure}[t!]
    \centering
    \includegraphics[width=0.95\textwidth]{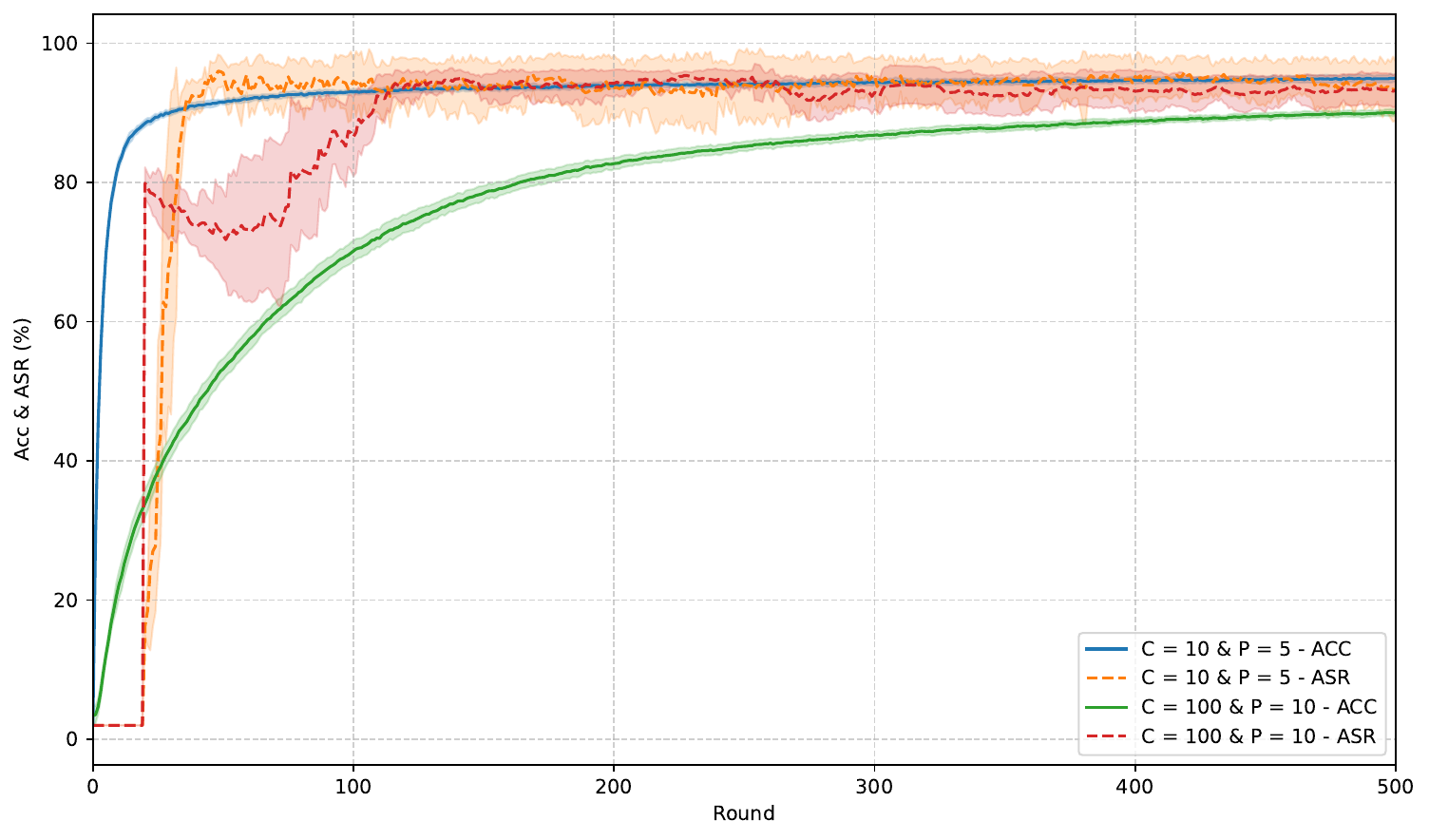}
    \caption{Influence of the client subset ($|\mathcal{P}|$) and the total number of clients ($|\mathcal{C}|$).}
    \label{fig:various_clients_resnet18_gtsrb}
\end{figure}

FL systems, particularly in cross-device scenarios, often involve highly heterogeneous client populations, complicating the evaluation of attacks and the establishment of benchmarks. The two primary parameters are (1) the size of the client subset ($|\mathcal{P}|$), which affects the dilution of the aggregation, and (2) the total number of clients ($|\mathcal{C}|$) which determines how frequently malicious clients appear. We evaluate the scalability of our attack by increasing the client count to $C=100$ and $P=10$. Fig.~\ref{fig:various_clients_resnet18_gtsrb} presents the ACC and ASR evolution:  
the backdoor injection relying on CoBF succeeds even with a larger number of clients and a stronger dilution effect. In this setting, the ASR reaches values comparable to those obtained in the previous experiments in section~\ref{experiments_results}. Note that, for $|\mathcal{C}|=100$ and $|\mathcal{P}|=10$, the ASR starts at $80\%$ because of the low training convergence and the weak performance of the model ($\approx30\%$). However, with Table \ref{tab_nb_faults_cobf_fl_setting}, we observe that, with more clients, the fault count increases drastically from 165 to 347 faults and slightly more than twice as many participations are required to successfully perform the backdoor injection (44 vs. 19).

\input{tab/stats_faults_rounds_ResNet18}

\subsection{Mitigation}
\label{mitigation}
 
In the FL context, defenses against backdoors are all focused on data-based poisoning approach and propose to restrict or reject \textit{abnormal} client updates during the aggregation, such as the standard Norm Clipping \cite{sun2019can} that we analyze hereafter. Note that, all the defenses based on filtering training data to detect salient triggers are ineffective against parameter-based backdoor attacks. Similarly, proposed defenses against hardware-based backdoor attacks primarily target inference-time attacks by analyzing models with static parameters \cite{chen2021proflip,rakin2020tbt} and are not relevant for attacks at training-time. In our case, defending against the CoBF attack can be achieved by exploiting two inherent weaknesses: (1) the requirement for relatively stable parameters around the PTM values; (2) the tendency of bit-flips to increase the absolute value of parameters, usually pushing them outside their standard distribution. Consequently, we present two studies focusing on the influence of the learning rate and on Norm Clipping.

\input{tab/isr_LR_NFPA}

\paragraph{Adapting the learning rate.} During the training, the learning rate ($\lambda$) is one of the hyperparameters that heavily influences the optimization and, then, the potential stability. We investigate its impact on the attack's effectiveness in Table \ref{table_lr_nfpa_variation_resnet18_gtsrb} with the \textit{injection success rate}, i.e. the percentage of faults identified with the CoBF method that are successfully applied online, during the model adaptation. This analysis is based on different NFPA values, where a lower NFPA corresponds to a larger attack window, resulting in faults being injected further into the training process (and thus further from the PTM).

We observe that $\lambda$ and NFPA affect the effectiveness of CoBF. Across all NFPA values, decreasing $\lambda$~--~thereby increasing model stability~--~correlates with a higher injection success rate. Conversely, with a $\lambda$=1e-2, the attack becomes highly unstable across runs and most often fails to inject the backdoor. Furthermore, the induced faults disrupt the training of the benign task, making the attack detectable. 
At $\lambda$=1e-3 with NFPA=10, we achieve a near-perfect fault injection success rate ($97.6\%$), which drops by 9.7\% simply by switching to $\lambda$=5e-3. Moreover, across all learning rate values, we observe that a shorter attack windows (NFPA=10)~--~i.e., more faults per apparition of the malicious client~--~correlates with a higher injection success rate. Thus, for a minimal adversarial budget (NFPA=1), $\lambda$ plays a significant influence: by increasing it to 5e-3, the injection success rate drops to 62.6\%, which substantially slows down the lasting establishment of the backdoor. 

\begin{figure}[t!]
    \centering
    \includegraphics[width=0.95\textwidth]{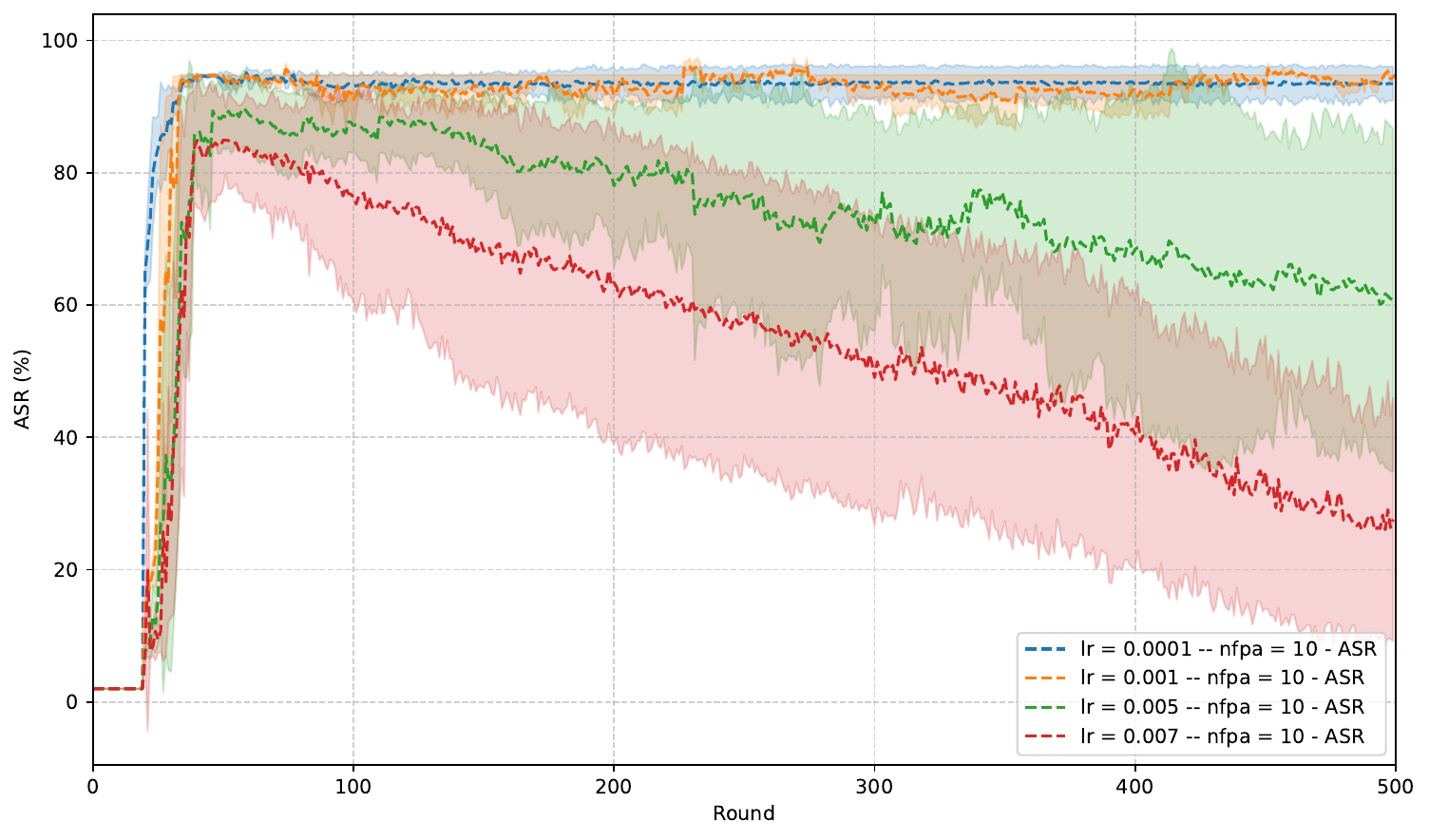}
    \caption{Influence of the learning rate on the ASR (ResNet-18 on GTSRB).}
    \label{fig:various_lr_resnet18_GTSRB}
\end{figure}

The influence of $\lambda$ over the ASR is reflected in Fig.~\ref{fig:various_lr_resnet18_GTSRB}
. Here, relative to our experimental baseline of 1e-3, we use one lower value (1e-4) and two higher ones (5e-3, 7e-3). Importantly, starting from 7e-3, we observed that both the attack and nominal task learning became unstable across different runs, rendering the attack ineffective as it becomes detectable. Lowest values (1e-3, 1e-4) lead to the strongest and most stable and persistent backdoors (94\% after 500 rounds). On the contrary, for higher learning rates, the attack's variability across different runs increases significantly (whereas little variance is observed at 1e-3), and the backdoor is erased much more rapidly following the injection phase.

From this analysis, we conclude that a best practice for defending against this attack is to select the highest learning rate among the most suitable and efficient values for the targeted FL system.

\paragraph{Robustness of CoBF against Norm Clipping. }

The main limitation of fault injection attacks against parameters is the risk of pushing the faulted values outside the nominal distribution of their original layer. This is identified in the centralized attack DeepVenom \cite{cai2024deepvenom} as a potential defense mechanism by bounding parameter values~--~though the authors note that an attacker can adapt the fault magnitude, albeit at the cost of requiring significantly more injections. In FL, this corresponds to the standard defense against backdoor attacks known as Norm Clipping (NC) \cite{sun2019can,zhang2023a3fl,zhang2022neurotoxin}. Basically, at round $(t)$, NC clips the local parameters $W_c^{(t)}$ of a client $c\in \mathcal{P}$, w.r.t. a threshold $\tau$ (Eq. \ref{eq_nc}). Standard value are $\tau \geq 1$ as in state-of-the-art backdoor attacks evaluation \cite{zhang2023a3fl,zhang2022neurotoxin}.

\begin{equation}
    \text{Clip}(W_c^{(t)},\tau)=W_c^{(t)}/max\big( 1,||W_c^{(t)}||_2/\tau\big)
    \label{eq_nc}
\end{equation}

\begin{figure}[t!]
    \centering
    \includegraphics[width=\textwidth]{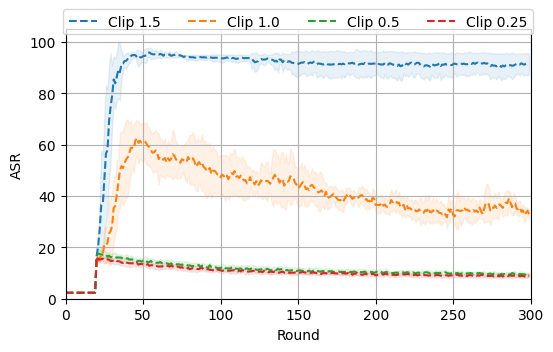}
    \caption{Impact of Norm Clipping (NC)}
    \label{fig:CoBF_against_clipping}
\end{figure}

Fig. \ref{fig:CoBF_against_clipping} shows the impact of NC against our attack (ResNet-18, GTSRB). If $\tau$ is not well suited, the backdoor is successful (or partially successful with $\tau=1$). However with lower values of clipping, faults impact is too significantly truncated and the backdoor is not properly injected. 

\begin{figure}[t!]
    \centering
    \includegraphics[width=\textwidth]{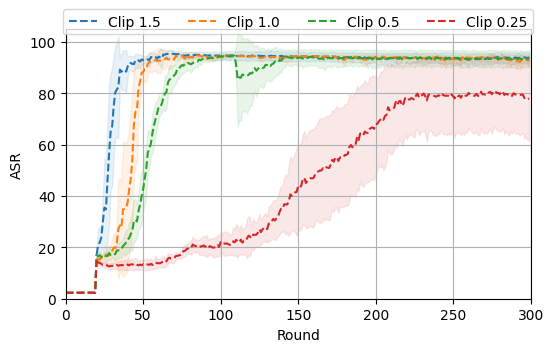}
    \caption{Adaptive CoBF with multiple bit-flips against NC}
    \label{fig:multiflip_against_clipping}
\end{figure}

Knowing the NC baseline and guessing relevant values for $\tau$, an adaptive attacker may design a new version of CoBF by reducing the impact of the faults in a single round to avoid being clipped. While the difference between a parameter and its faulted version exceeds $\tau$, the attacker will apply new faults in the same round to reduce the impact on the weight and 
induce a slower and stealthier influence on the global model.  This strategy is straightforward to derive from Algo \ref{alg:cap}, and its implementation is proposed in the Appendix (Algorithm \ref{alg:cap_multi}). In Fig. \ref{fig:multiflip_against_clipping}, we see that this adaptive CoBF algorithm significantly improves the ability to circumvent NC, with the exception of the lowest threshold ($\tau=0.25$) for which the injection time is longer (with a plateau at 80\%). Nevertheless, the effectiveness of this new attack relies on a significant increase in the number of faults (on average, $\times4$ more faults)~--~leading to longer injection times~--~and on the execution of multiple bit-flips on the same parameter within a single round. Thereby, it violates the core hypothesis of Rowhammer viability, since the primary assumption is that only bits located in different memory pages are targeted \cite{tol2023don}. Consequently, while the attack is viable from a purely algorithmic perspective, its practical implementation via injection mechanisms such as Rowhammer remains unrealistic. There are likely other optimization strategies that remain to be explored in future works. This initial analysis of an adaptive attack paves the way for future works into model poisoning optimization through the fine-grained manipulation of parameters.

\subsection{Attack practicality}
Our results are derived from simulations and aim to characterize and evaluate the impact of a fault-based attack vector in a federated context. Future work will require investigating the actual (and, in FL, complex) practical implementation of this attack across hardware platforms and model architectures using a Rowhammer-type injection method. However, following the recommendations in \cite{tol2023don}, we have applied several constraints that are compatible with Rowhammer. Regarding the practicality of the attack, we highlight the fact that  attacking within a distributed paradigm, where the impact of faults accumulates round after round, provides a distinct advantage to the adversary. Indeed, our attack relies on a limited number of faults (NFPA) each time the malicious client is selected, and the attacker can leverage the time between two participations to optimize their injection (weight matrix reallocation in memory in case of Rowhammer attacks).

%% file: tab/stats_faults_rounds_ResNet18.tex
\begin{table}[t!]
\caption{Faults and injection rounds for ResNet-18 with two FL settings}
\label{tab_nb_faults_cobf_fl_setting}
\centering
\begin{tabular}{lcccc}
\toprule
 & \# Faults in COFF\_list  & \# Injection rounds & NF & NFPA \\
\midrule
{$|\mathcal{C}|=10$, $|\mathcal{P}|=5$} & 36 & 19 & 165 & 10\\
{$|\mathcal{C}|=100$, $|\mathcal{P}|=10$} & 36 & 44 & 347 & 10\\
\bottomrule
\end{tabular}
\end{table}

%% file: tab/isr_LR_NFPA.tex
\begin{table}[b!]
\caption{Impact of learning rate and NFPA on the fault injection success rate}
\centering
\begin{tabularx}{0.99\textwidth} { 
    >{\raggedright\arraybackslash}l     
    >{\centering\arraybackslash}X  
    >{\centering\arraybackslash}X  
    >{\centering\arraybackslash}X  
    >{\centering\arraybackslash}X  
    >{\centering\arraybackslash}X  
   }
\toprule
 & \multicolumn{5}{c}{NFPA} \\
$\lambda$ & 1 & 2 & 3 & 5 & 10 \\
\midrule
$10^{-4}$& $97.4 \pm 1.3$    & $97.9 \pm 0.3$    & $98.2 \pm 0.0$    &  $97.3 \pm 0.9$   & $98.2 \pm 0.0$    \\
$10^{-3}$& $84.3 \pm 1.7$ & $93.5 \pm 2.0$ & $94.1 \pm 1.1$ & $97.6 \pm 0.5$ & $97.6 \pm 0.8$ \\
$5.10^{-3}$& $62.6 \pm 1.4$ & $70.8 \pm 3.6$ & $77.9 \pm 1.3$ & $87.8 \pm 3.1$ & $87.9 \pm 3.3$ \\
$10^{-2}$& \multicolumn{5}{c}{unstable attack / attack failure}  \\
\bottomrule
\end{tabularx}
\label{table_lr_nfpa_variation_resnet18_gtsrb}
\end{table}

%% file: src/conclusion.tex
This paper introduces the first backdoor technique for FL systems leveraging HW-based fault injection attacks during the local training of a client. As in a centralized context, with demonstration by Rowhammer exploits on DRAM memory, model poisoning in FL represents a powerful attack vector that operates under distinct threat models compared to classical data-based poisoning approaches. We demonstrate that the primary barrier intrinsic to FL, the dilution of faults caused by the clients aggregation, can be effectively bypassed. To this end, we introduce the Chain-of-Bit-Flip (CoBF) principle, which 
estimates the contributions of other clients and schedules the attack through a succession of intermediate faults. These faults progressively steer specific local model parameters~--~identified during an offline phase on the PTM with the WBP centralized mechanism~--~toward the desired target values. Our results confirm the feasibility of installing a backdoor  
on state-of-the-art models. 
Notably, CoBF enables the installation of a highly persistent backdoor on a ResNet-18 with only 19 malicious participations of the targeted client and a maximum of 10 bit-flips per round. Furthermore, we outline best practices to mitigate this threat, such as maintaining a sufficiently high learning rate or deploying Norm Clipping with suitable clipping threshold.  
While it remains theoretically possible to bypass such defense by adapting CoBF, doing so would require an attacker to perform a significantly higher volume of faults at a granularity that would likely render the attack infeasible via Rowhammer-based injection.

The main perspectives and future directions involve a deeper analysis of the attack’s mechanisms across varying model architectures and tasks (as discussed in \ref{experiments_results}), including backdoor tasks in NLP and LLMs. To the best of our knowledge, this work is the first to investigate physical attacks with fault injection against the integrity of FL systems. It paves the way for necessary studies on practical exploration, ranging from diverse injection vectors (e.g., advanced Rowhammer variants, EM pulse or laser injection) and alternative fault models (e.g., data-dependent faults or instruction skip) to the efficiency of classical fault-tolerance methods (e.g., redundancy, Error Correction Codes). These research avenues are fundamental as FL increasingly migrates toward critical systems and as future regulatory and certification frameworks for these systems emerge.

%% file: src/appendix.tex
\noindent\textbf{Benign performance.} Table \ref{acc_tab} gives the average performance (ACC, over 5 runs) of each model on their respective downstream tasks. The accuracy is measured at round 100 (same setup as for experiments in Fig \ref{fig:asr_all_models}).  

\begin{table}[h!]
\caption{Accuracy of the models for the benign tasks.}
\label{acc_tab}
\centering
\begin{tabular}{lccc}
\toprule
&CIFAR-10&EuroSat&GTSRB\\
\midrule
\textbf{ResNet-18} & 87.54 (0.16) & 95.68 (0.68) & 92.90 (0.35)\\
\textbf{VGG-16} & 90.45 (0.25) & 96.68 (0.37) & 91.4 (0.62) \\
\toprule
& CIFAR-100 & Flowers-102 & Pets-37 \\
\midrule
\textbf{ViT} & 90.93 (0.1) & 99.18 (0.15) & 92.50 (0.25) \\
\bottomrule
\end{tabular}
\end{table}

\begin{figure}[h!]
    \centering
    \includegraphics[width=0.95\textwidth]{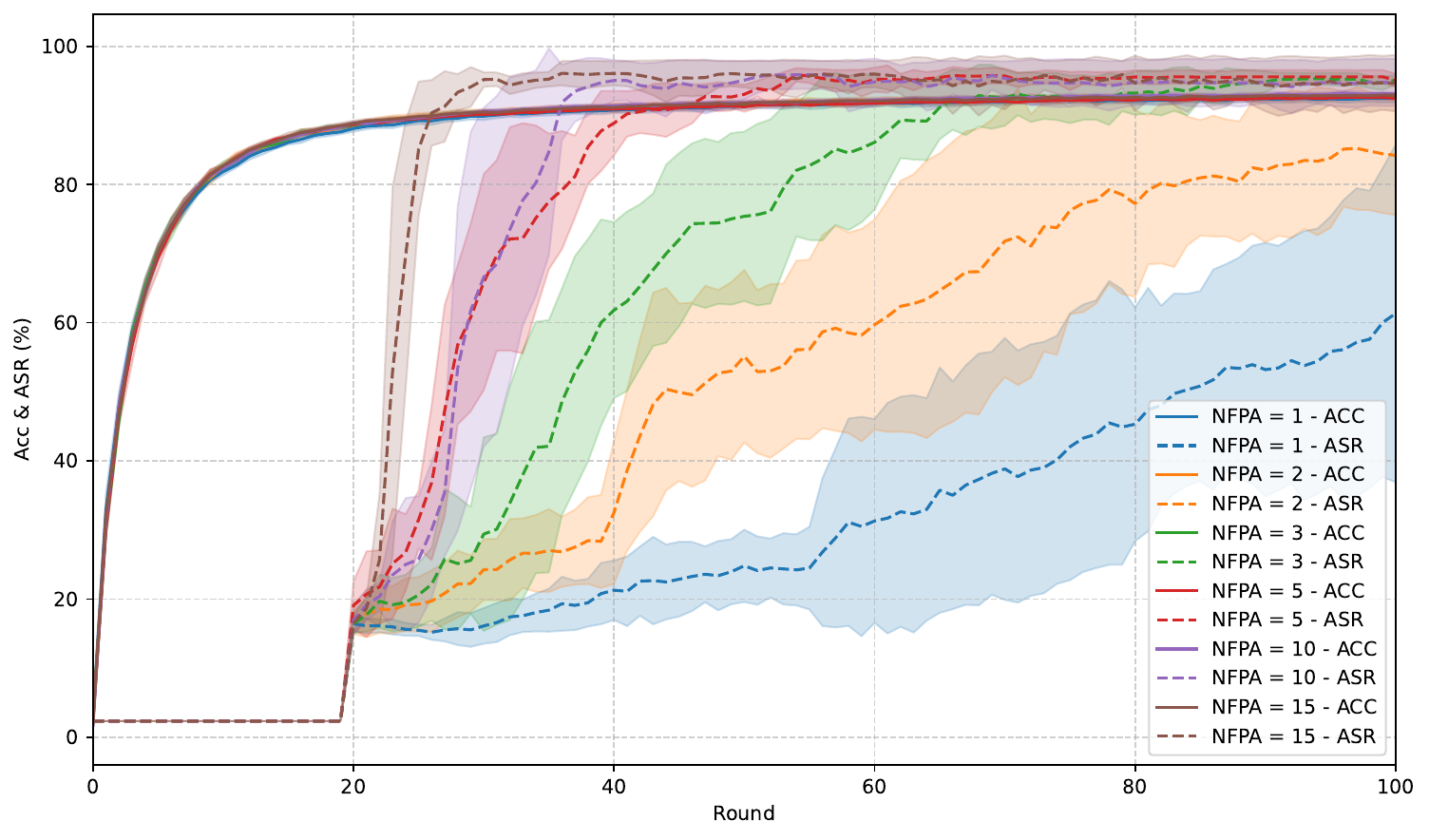}
    \caption{ACC and ASR evolution with different numbers of faults per appearance~--~NFPA (ResNet-18 on GTSRB).}
    \label{fig:zoom-in_nfpa}
\end{figure}

\noindent\textbf{Influence of NFPA on the backdoor injection.} Fig. \ref{fig:zoom-in_nfpa} shows the ASR and ACC evolution when using different NFPA for the ResNet-18 model (on GTSRB). The figure is focused on the begining of the injection step, before round 100. 

\noindent\textbf{Adaptive version of CoBF against Norm Clipping. } As described in Section \ref{mitigation}, CoBF can be mitigated by Norm Clipping which is one of the baseline defenses in FL. However, with the knowledge of the clipping level $\tau$, an adaptive attacker may improve the CoBF principle by injecting multi-flip fault instead of a single bit-flip. The change in the algorithm concerns the line 13 in Algorithm \ref{alg:cap} that we modify as follows:

\begin{algorithm}
$13: \text{multiflip},w^* \leftarrow \text{create-multiflip}(w^{(t)},w^*,p,i^{(t)},\tau)$ \\
$14: \text{bf-chain.append(multiflip)}$
\end{algorithm}

Where $\tau$ is the factor corresponding to the estimated clipping factor of the defense, the multi-flip will ensure having an impact on the parameter lower than $\tau$. The multi-flip function is described in Algorithm \ref{alg:cap_multi}.

\begin{algorithm}[h!]
\caption{$create\_multiflip$}\label{alg:cap_multi}
\begin{algorithmic}[1]
\Require $w, w^*, p, i, \tau$
\Ensure $multiflip$
\State $multiflip \gets [(p, i)]$
\While{$|w^* - w| > \tau$}
    \If{$w^*[i+1]$ is $1$}
        \State $w^*[i+1] \gets w^*[i+1]\oplus 1$
        \State $multiflip.append((p, i+1))$
    \ElsIf{$w^*[i+1]$ is $0$}
        \State $multiflip \gets []$
        \State $w^* \gets w$
        \State $w^*[i+1] \gets w^*[i+1] \oplus 1$
        \State $multiflip.append((p, i+1))$
    \EndIf
\State $i \gets i+1$
\EndWhile
\end{algorithmic}
\end{algorithm}

The \textit{multiflip} output is a list of bit-flips which has to be performed in the same round. As discussed in Section \ref{mitigation}, this algorithm raises critical practical issue when dealing with the hardware constraints for Rowhammer as proposed in \cite{tol2023don} (and applied in our CoBF experiments), since several bit-flips on the same parameter have to be executed during the same round.

%% file: bibliography.bib
@inproceedings{mcmahan2017communication,
  title={Communication-efficient learning of deep networks from decentralized data},
  author={McMahan, Brendan and Moore, Eider and Ramage, Daniel and Hampson, Seth and y Arcas, Blaise Aguera},
  booktitle={Artificial intelligence and statistics},
  pages={1273--1282},
  year={2017},
  organization={PMLR}
}

@inproceedings{chen2021proflip,
  title={Proflip: Targeted trojan attack with progressive bit flips},
  author={Chen, Huili and Fu, Cheng and Zhao, Jishen and Koushanfar, Farinaz},
  booktitle={Proceedings of the IEEE/CVF International Conference on Computer Vision},
  pages={7718--7727},
  year={2021}
}

@inproceedings{rakin2019bit,
  title={Bit-flip attack: Crushing neural network with progressive bit search},
  author={Rakin, Adnan Siraj and He, Zhezhi and Fan, Deliang},
  booktitle={Proceedings of the IEEE/CVF International Conference on Computer Vision},
  pages={1211--1220},
  year={2019}
}

@inproceedings{rakin2020tbt,
  title={Tbt: Targeted neural network attack with bit trojan},
  author={Rakin, Adnan Siraj and He, Zhezhi and Fan, Deliang},
  booktitle={Proceedings of the IEEE/CVF conference on computer vision and pattern recognition},
  pages={13198--13207},
  year={2020}
}

@inproceedings{tol2023don,
  title={Don't knock! rowhammer at the backdoor of dnn models},
  author={Tol, M Caner and Islam, Saad and Adiletta, Andrew J and Sunar, Berk and Zhang, Ziming},
  booktitle={2023 53rd Annual IEEE/IFIP International Conference on Dependable Systems and Networks (DSN)},
  pages={109--122},
  year={2023},
  organization={IEEE}
}

@inproceedings{zhang2022neurotoxin,
  title={Neurotoxin: Durable backdoors in federated learning},
  author={Zhang, Zhengming and Panda, Ashwinee and Song, Linyue and Yang, Yaoqing and Mahoney, Michael and Mittal, Prateek and Kannan, Ramchandran and Gonzalez, Joseph},
  booktitle={International Conference on Machine Learning},
  pages={26429--26446},
  year={2022},
  organization={PMLR}
}

@article{zhang2023a3fl,
  title={A3fl: Adversarially adaptive backdoor attacks to federated learning},
  author={Zhang, Hangfan and Jia, Jinyuan and Chen, Jinghui and Lin, Lu and Wu, Dinghao},
  journal={Advances in neural information processing systems},
  volume={36},
  pages={61213--61233},
  year={2023}
}

@inproceedings{xie2019dba,
  title={Dba: Distributed backdoor attacks against federated learning},
  author={Xie, Chulin and Huang, Keli and Chen, Pin-Yu and Li, Bo},
  booktitle={International conference on learning representations},
  year={2019}
}

@inproceedings{li2025rowhammer,
  title={Rowhammer-Based Trojan Injection: One Bit Flip Is Sufficient for Backdooring DNNs},
  author={Li, Xiang and Meng, Ying and Chen, Junming and Luo, Lannan and Zeng, Qiang},
  booktitle={34th USENIX Security Symposium (USENIX Security 25)},
  pages={6319--6337},
  year={2025}
}

@article{sun2019can,
  title={Can you really backdoor federated learning?},
  author={Sun, Ziteng and Kairouz, Peter and Suresh, Ananda Theertha and McMahan, H Brendan},
  journal={arXiv preprint arXiv:1911.07963},
  year={2019}
}

@article{wang2020attack,
  title={Attack of the tails: Yes, you really can backdoor federated learning},
  author={Wang, Hongyi and Sreenivasan, Kartik and Rajput, Shashank and Vishwakarma, Harit and Agarwal, Saurabh and Sohn, Jy-yong and Lee, Kangwook and Papailiopoulos, Dimitris},
  journal={Advances in neural information processing systems},
  volume={33},
  pages={16070--16084},
  year={2020}
}

@inproceedings{cai2024deepvenom,
  title={Deepvenom: Persistent dnn backdoors exploiting transient weight perturbations in memories},
  author={Cai, Kunbei and Chowdhuryy, Md Hafizul Islam and Zhang, Zhenkai and Yao, Fan},
  booktitle={2024 IEEE Symposium on Security and Privacy (SP)},
  pages={2067--2085},
  year={2024},
  organization={IEEE}
}

@inproceedings{cai2024wbp,
  title={Wbp: Training-time backdoor attacks through hardware-based weight bit poisoning},
  author={Cai, Kunbei and Zhang, Zhenkai and Lou, Qian and Yao, Fan},
  booktitle={European Conference on Computer Vision},
  pages={179--197},
  year={2024},
  organization={Springer}
}

@inproceedings{dong2023one,
  title={One-bit flip is all you need: When bit-flip attack meets model training},
  author={Dong, Jianshuo and Qiu, Han and Li, Yiming and Zhang, Tianwei and Li, Yuanjie and Lai, Zeqi and Zhang, Chao and Xia, Shu-Tao},
  booktitle={Proceedings of the IEEE/CVF International Conference on Computer Vision},
  pages={4688--4698},
  year={2023}
}
